\documentclass{eas}
\usepackage{graphicx}
\begin{document}

\title{Dense Star-forming Gas and Dust in the Magellanic Clouds} 
\author{F.P. Israel}\address{Sterrewacht Leiden, Leiden University, Postbus 9513, 2300 RA Leiden, the Netherlands}
%
%
\begin{abstract}
The early stages of the star formation process are closely related to
the condition of the parent interstellar medium, in particular to the
heating/cooling balance, which itself is a function of ambient
conditions. Important questions such as dust abundance, size
distribution, temperature distribution, fraction of molecular gas,
fraction of dense gas, gas surface density and total amount of gas and
dust require separation of metallicity and radiation effects. The
Magellanic Clouds provide an ideal laboratory to carry out such
studies almost under 'controlled conditions'.  Although they are
prominent targets for space observatories (Spitzer, Herschel), an
important role remains for large groundbased facilities, such as a
25 m class sub-millimeter telescope on Dome C.  Large-scale mapping at
high resolution should be carried out both in the continuum and in
various lines, fully complementing other groundbased and spacebased
observing programs.
\end{abstract}
\maketitle
\section{Introduction}

The formation of stars out of the tenuous interstellar medium is a
process involving condensation of gas, mostly hydrogen, by a factor of
the order of 10$^{23}$. The beginning of this process is usually
discussed in mostly qualitative terms, and even less is known about
how it proceeds in the early stages. In particular, it is not obvious
how the initial contraction proceeds, how contracting cores subdivide,
what determines how many stars are formed, and what will be the
distribution of their masses (see e.g. Myers 2005, 2008). Once stars
are formed, and nuclear fusion provides a huge increase in energy
output lighting up the star as well as its environments, the situation
becomes much more accessible to observers. However, the very energy
output that draws our attention to newly formed stars also overwhelms
the star's surroundings, rapidly changing their physical condition and
wiping out any clues to the initial conditions that existed before the
star was born.

Although there may be various ways to start this process, all assume
that the early stages involve the formation and evolution of clouds of
molecular gas and dust, having astrophysically high densities of at
least a thousand up to a million atoms per cc. In our Milky Way,
research is focusing on nearby quiescent dark clouds and dense cloud
cores to elucidate the early stages of star formation stages by
looking at the physical conditions in these dense clouds of gas and
dust. Of great significance is the heating and cooling balance as a
function of cloud properties such as size, volume density and column
density, temperature, composition etc. The clouds are not isolated
entities but are especially sensitive to ambient conditions, because
most clouds have filamentary and fractal structures, which means that
they are mostly surface and little volume.  Molecular gas and dust are
subject to processing as a function of irradiation (energy input) on
the one hand and shielding on the other hand. Depending on the degree
of processing, the properties of the molecular gas and the dust may
undergo considerable change.  It is important to be aware that
interstellar clouds of any type are not static in nature, but instead
are highly dynamical entities.

Major questions regarding interstellar dust are: how much dust is
there?  what is the dust particle composition?  what is the grain size
distribution?  Perhaps surprisingly, the answers to these questions
are still debated, even for nearby dust in the Solar Neighbourhood.
The answers depend on the outcome of a complex interplay of
interstellar processes and conditions controlling accretion, erosion
and even destruction of individual dust grains.

Major questions on dense gas are: how much molecular gas is there?
What is the fraction of very dense gas? what is the fraction of
well-shielded gas? Molecular gas clouds consist almost entirely of
molecular hydrogen (H$_{2}$).  By its symmetric structure, molecular
hydrogen is very difficult to observe, especially at the low
temperatures (10 - 100 K) characterizing the quiescent interstellar
medium. It has no convenient bi-pole transitions. When cold, it is
observable only in the UV, in absorption against a bright background.
At near-and mid-infrared wavelengths it can be observed directly, but
only in an excited condition. In either case, only tiny and
unrepresentative fractions of all molecular gas are sampled. Thus, we
can only study the bulk of molecular gas clouds by measuring its
tracers. This raises questions such as: what tracer to use? how
reliable are the answers provided by these tracers? The answers depend
on detailed knowledge of the ambient chemistry, in turn dependent on
elemental abundances, shielding and irradiation, in particular the
conditions governing the process of molecular photo-dissociation.

\section{Dust: an elusive ISM component}

Dust particles are an important component of the interstellar medium.
They shield fragile molecules such as CO from destruction by UV
photons, and they allow other molecules such as H$_{2}$ to form by
grain-surface reactions. Dust grain populations depend strongly on
local conditions, such as metallicity and energy density.  For
instance, shocks from expanding SNRs are a major and efficient source
of dust particle erosion and destruction on short timescales (Jones et
al. 1994, 1996). This rapid erosion may be countered by efficient
accretion of dust particles in quiescent, dense (molecular) gas
clouds.  Several authors have attempted to extract the resulting
equilibrium particle size distribution from measurements of
Solar-Neighbourhood objects (e.g. Mathis et al 1977).  At the small
end of their distribution, hot and very small grains excited by the
absorption of a single photon (Draine $\&$ Anderson, 1985) and even
smaller PAHs (L\'eger $\&$ Puget 1984) occur.  Including these
additional populations, Weingartner $\&$ Draine (2001) constructed
metallicity-dependent size distributions for carbonaceous and silicate
grains pertinent to the different conditions in the Galaxy, the LMC
and the SMC and other physically-based dust emission models are also
becoming available (e.g. Li $\&$ Draine 2001; Zubko et al. 2003),
 
As the erosion timescales are short compared to accretion timescales,
variations in ambient conditions may rapidly and substantially alter
local grain size distributions.  Alteration of the balance between big
grains and very small grains may cause noticeable changes in SED
shapes.  This appears to be the case in the strong star-burst
environment characterizing the dwarf galaxy NGC 1569 (Lisenfeld et
al. 2002).  A small but clear emission excess occurs on the
Rayleigh-Jeans side of the far-infrared peak. It is unlikely to be
caused by cold `classical' big grains.  It is much more probable that
this `excess' originates in very small grains completely out of
thermal equilibrium, with a relatively small total mass.  Thus, in
environments with extensive dust grain processing, traditional
interpretations easily overestimate dust masses, and consequently
underestimate both gas-to-dust ratios and gas masses deduced from dust
emission.

As a major part of the astration cycle, the ISM plays an essential
role in galaxy evolution.  It is thus important to gain insight in the
processes that govern dust grain processing and dust particle size
distributions.  Studying far-infrared/sub-millimeter SEDs as a function
of varying ambient conditions is a very promising way of making
significant progress.

\section{Molecular gas: a controversial ISM component}

There is no doubt that most late type galaxies - both spiral and
irregular galaxies - contain significant amounts of molecular gas.
However, is this true for {\it all} late type galaxies, in particular
the dwarf irregulars with the lowest metallicities (cf. Leroy et
al. 2007)? An even more importantly: how much molecular gas is there?
As mentioned before, the bulk of the molecular gas cannot be observed
directly, and must be revealed by tracer observations. Here the
controversy starts. Traditionally, the second most abundant molecule,
carbon monoxide (CO), was used to trace H$_{2}$.  CO column densities
were calculated from the measured intensities of the optically thick
$^{12}$CO line and the optically thin ${13}$CO line, assuming LTE
conditions. Molecular masses could then be determined by applying the
appropriate abundance ratio.  Alternatively, nearby Galactic molecular
clouds were mapped in CO, and by assuming the clouds to be in virial
equilibrium, masses could be determined.  Finally, a rough estimate
resulted from an analysis of low-energy gamma-ray emission (Bloemen et
al.  1986). All three methods seemed to yield consistent results,
implying a remarkably constant ratio between the optically thick CO
line luminosity, and the H$_{2}$ column density. This standard
CO-to-H$_{2}$ conversion factor became known as the $X$-factor.  It
is, however, on shaky ground as LTE conditions do not apply to most of
the CO-emitting gas, and the structure of molecular clouds as well as
their limited lifetimes do not justify use of the virial
theorem. Several authors have expressed doubt as to the validity of a
'standard' $X$-factor, using e.g. the far-infrared or sub-millimeter
emission from dust as a tracer to finding values up to an order of
magnitude different.  A more accurate and detailed gamma-ray analysis
led Strong et al. (2004) to suggest a significant radial variation
across the Milky Way. Ignoring these the `standard' factor was and
still is by some conveniently applied to CO observations of other
galaxies, no matter whether we are dealing with a poorly shielded
dwarf galaxy, low in carbon content, or with a very luminous young
galaxy at high cosmological redshifts. However, use of other tracers
than CO is not without its own problems.  Dust emits as a very strong
function of temperature. This introduces a significant uncertainty in
column density, as does the uncertainties in dust emissivity and
abundance.  Another tracer is ionized carbon, but it does not
distinguish between molecular and atomic hydrogen, and the derived
molecular gas column densities depend on carbon chemistry, abundance,
and excitation.

In order to derive reliable molecular gas column densities the
problems associated with their tracers need to be resolved. This
requires better and detailed determination of CO physics and
photodissocation, better and detailed determination of dust properties
and temperatures, and better and detailed observation of ionized
carbon, and preferably also neutral carbon in order to completely
determine the carbon excitation and abundance. 

\begin{figure*}
\begin{center}
\begin{minipage}{8.cm}
\resizebox{8.cm}{!}{\rotatebox{0}{\includegraphics*{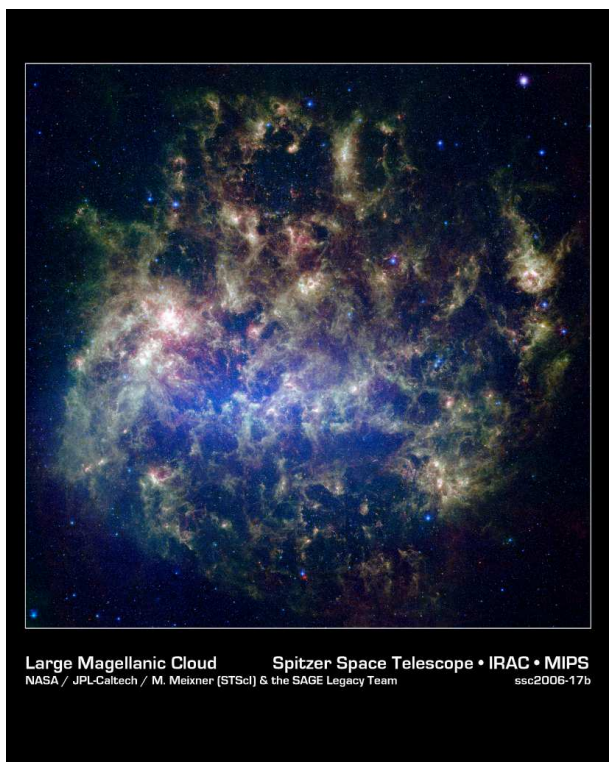}}}
\end{minipage}
\begin{minipage}{3.cm}
\resizebox{3.cm}{!}{\rotatebox{0}{\includegraphics*{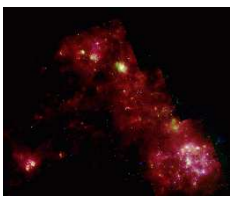}}}
\end{minipage}
\caption{Infrared images of the LMC (left) and the SMC (right) from
  Spitzer Space Observatory data.  Images produced by the SAGE and
  SMCSAGE collaborations.  }
\end{center}
\end{figure*}

\section{The Magellanic Clouds: ideal `controlled environment' laboratories}

Further progress in the analysis of the dense interstellar medium in
general requires accurate determination of the influence of the
environment under varying conditions. In physics, experiments may be
conducted by systematically changing one variable after another. In
astronomy, only the 'experiments' presented by nature are available.
In case of the ISM, much would already be gained if we could analyze
its properties systematically as a function of ambient radiation field
and metallicity (shielding). This requires a set of environments of
{\it distinctly different metallicity}, each environment of a {\it
  constant metallicity throughout}, and each environment {\it
  presenting a range of radiation field strengths}. These environments
should be observable with a good linear resolution so that each
resolution element in principle samples a homogeneous section of the
environment, eliminating as much as possible confusion by
beam-smearing. The baseline environment is the Solar Neighbourhood in
the Milky Way galaxy. The close Milky Way satellites, the Large
Magellanic Cloud (LMC) and the Small Magellanic Cloud (SMC) provide
the ideal astrophysical laboratories for the complementary
investigations.  With distances of 50 kpc (Feast 1999) and 61 kpc
(Keller $\&$ Wood 2006) respectively, they are ten times closer than
other major systems in the Local Group of galaxies, and typically
40-100 times closer than other `nearby' galaxies.  Their external
nature but close proximity simultaneously provides full global and
unimpeded very detailed views. They allow us to resolve parsec-sized
structures, making it possible to study ISM processes unambiguously on
scales dominated by a single star.  They are gas-rich but have low
dust and metal abundances, different from each other and from the
Milky Way, thus providing three different metallicity environments.

The LMC, with a metallicity one quarter solar (Dufour et al.  1984),
has a mass of $9\,\times\,10^{9}$ M$_{\odot}$ (van der Marel et al.
2002), a gas fraction of one third, a size of about 8 kpc and a star
formation rate of 0.1 M$_{\odot}$/yr (Whitney et al.  2007).  The SMC,
with a metallicity one tenth solar (Dufour et al.  1984) has a mass of
$2.5\,\times\,10^{9}$ M$_{\odot}$, a gas fraction of about a quarter,
a size of about 3 kpc, and a star formation rate of 0.05
M$_{\odot}$/yr (Wilke et al.  2003).  Metallicities show little
variation across either system, but luminous star formation has
created a very large range of ambient radiation field intensities in
each Cloud.  In the LMC, over 300 ionized gas complexes associated
with various degrees of star formation have been identified; in the
SMC over half that number. UV radiation field intensities in both
Clouds range from less than those in the Solar Neighbourhood to
hundreds of times higher.

This allows us to study the effects of ambient radiation fields at
constant metallicities.  Because of their particular nature, the
Magellanic Clouds provide well-determined environments representing a
variety of metallicities and energy densities in which the resulting
properties of the interstellar medium can be studied in a detailed and
systematic manner not offered elsewhere.  Not only is by far the
largest part of the Milky Way inaccessible to such studies, but much
of the Milky Way suffers from confusion caused by line-of-sight
crowding and distance ambiguities.  In the LMC and the SMC, everything
is essentially at the same distance, and determinations of mass and
luminosity are unambiguous and directly comparable.  The LMC is
particularly favourable because its limited depth and its viewing
angle of $35^{o}$ (van der Marel $\&$ Cioni 2001) place only a single
cloud along any line of sight.

\section{Gas and dust processing in the Magellanic Clouds}

In low-metallicity dwarf galaxies, such as the LMC and the SMC, use of
tracers other than CO first led to different results.  Progress in
far-infrared observing techniques made it possible to use dust
continuum emission (NGC~6822, LMC, SMC, see Israel 1997) and the line
emission from ionized carbon [CII] (IC~10, cf. Madden et al. 1997) as
a tracer for the total amount of molecular gas. Although these tracers
also have some problems, the $X$-factors implied by them are so much
higher, by one to two orders of magnitude, than those found with the
traditional CO methods that they are hard to ignore. Most recent
studies confirm a dependence of the CO-to-H$_{2}$ ratio on metallicity
(and radiation intensity) but disagree on the measure. It appears that
studies encompassing large sections of the LMC and the SMC (`big
beams') yield significantly higher $X$-values than those concentrating
on individual CO clouds (cf. Israel 2001).  This is, however, easily
understood if dense central regions, containing H$_{2}$ and CO, are
embedded in less dense regions of molecular gas with CO largely
photo-dissociated into atomic carbon and oxygen, e.g. lacking CO
emission . For a recent investigation, discussing in detail the
problems associated with a reliable determination of molecular gas in
a low-metallicity SMC molecular cloud complex, see the study by Leroy
et al. (2009), which also contains many references to earlier work.
In the end, however, the most and perhaps the only reliable way to
establish molecular gas parameters is by careful and detailed
modelling of molecular line spectroscopy, for instance involving
several transitions of various isotopes of CO in particular, but also
higher-density tracers such as HCN, HCO$^{+}$, HNC, CS etc..

In both the LMC and the SMC, molecular gas is significantly processed,
changing the relation between molecular hydrogen and less abundant
species, starting with CO. The occurrence and importance of molecular
gas processing, i.c. CO photo-dissociation, is well-illustrated by a
comparison of the distribution of CO and ionized carbon [CII] emission
across the LMC (Figure 1). On large linear scales, CO and [CII]
emitting regions occur together, but on smaller scales the emission is
strongly anti-correlated.

\begin{figure*}
\begin{center}
\begin{minipage}{5.5cm}
\resizebox{5.5cm}{!}{\rotatebox{0}{\includegraphics*{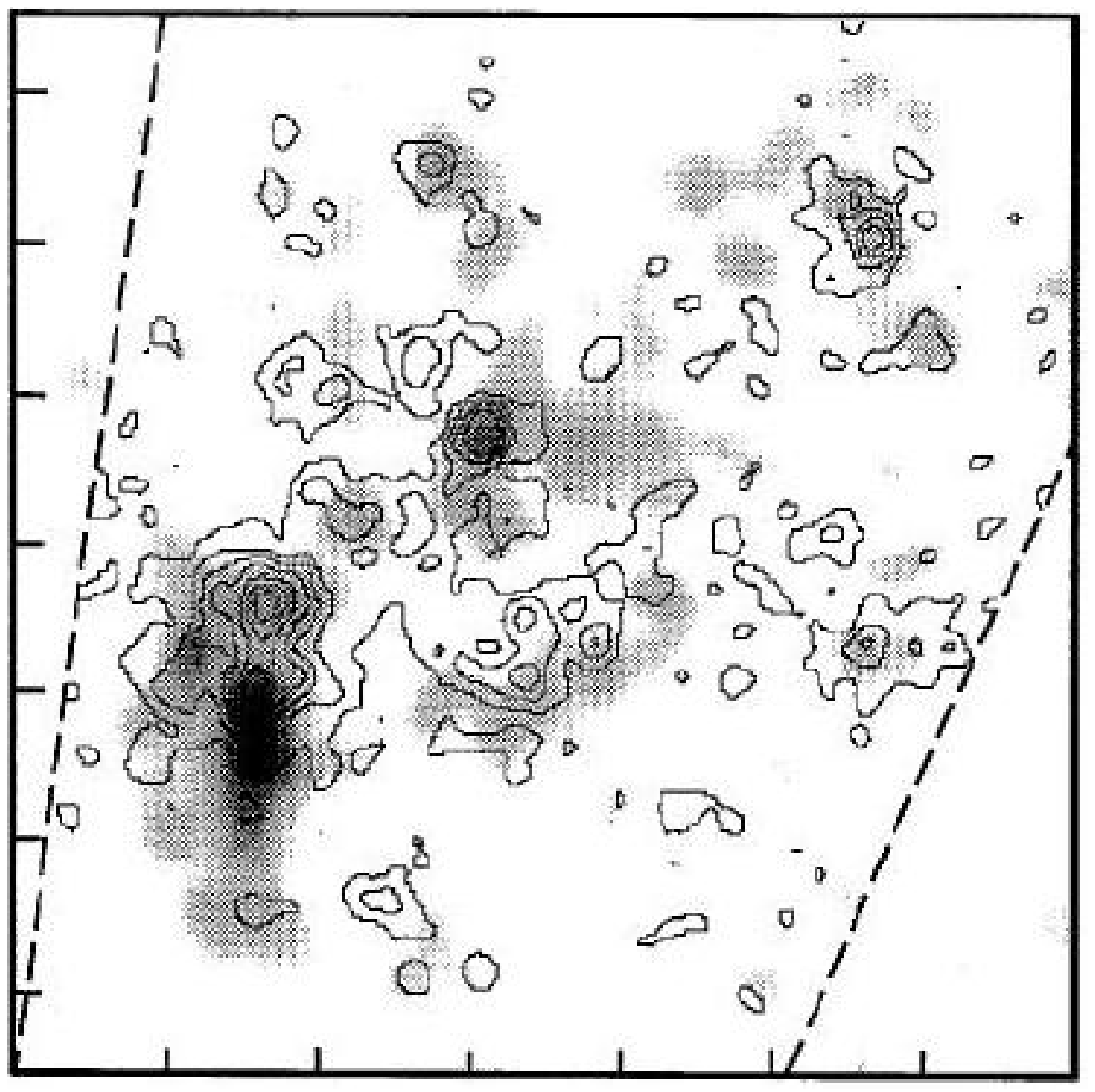}}}
\end{minipage}
\begin{minipage}{5.5cm}
\resizebox{5.5cm}{!}{\rotatebox{270}{\includegraphics*{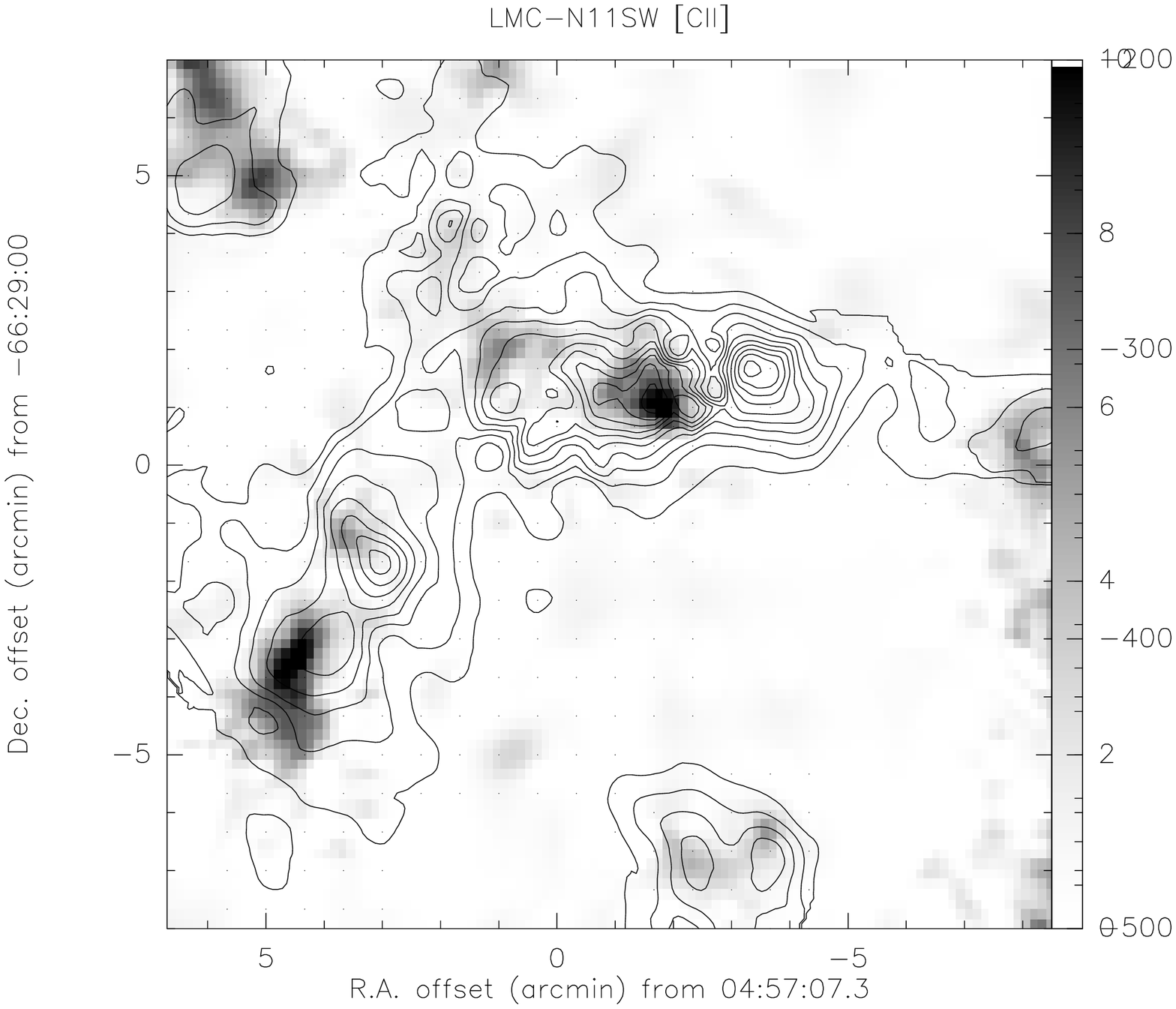}}}
\end{minipage}
\caption{The effects of CO photodissociation are very well illustrated
  by superposing ionized carbon ([CII]) contours on greyscale images
  of $J$=1-0 CO.  Left: all of the LMC in low resolution (Mochizuki et
  al. 1994). Right: part of the N~11 complex in the upper NW corner of
  the LMC (Israel et al., 2009). In both images, the associative but complemantary nature of
  the two is obvious.}
\end{center}
\end{figure*}

In both Clouds, dust is also processed, leading to changes in the
emission across the far-infrared/submillimeter part of the spectrum.
Not only the spectrum of NGC~1569 shows evidence for enhanced emission
in the Rayleigh-Jeans (RJ) wing of the infrared peak, such excess
emission has also been noted in the continuum flux distributions of
Virgo-cluster blue compact dwarf galaxies (Popescu et al.  2002), in
the late-type spiral star-burst galaxy NGC~4631 (Dumke et al. 2004), in
the Sombrero galaxy NGC~4594 (Bendo et al. 2005), and in the Milky
Way galaxy itself (Reach et al.  1995). A difficulty in interpreting
the local RJ excess emission is our lack of absolute sub-mm
measurements within the Milky Way, and an even greater difficulty in
extragalactic studies is the lack of spatial resolution and hence wide
range of temperatures within a single beam.  In the case of external
galaxies, the poor sampling of the RJ wing of the spectral energy
distributions and limited accuracy of the few data points available
considerably adds to this difficulty.

In the LMC and the SMC, the situation is more favourable. In a recent
study, Israel et al. (2009) present unusually well-sampled {\it
  global} spectral energy distributions of the Magellanic Clouds,
covering the RJ wing in the 30-1000 GHz (1 cm to 300 $\mu$m
wavelength) range. The RJ excess is clearly discernible in the LMC
spectrum, and even more pronounced in the SMC spectrum. No such strong
RJ excess occurs in the predominantly thermal spectra of larger
star-burst galaxies such as NGC~253, M~82, and NGC~4945.  In principle,
a number of different explanations may apply: large amounts of very
cold classical dust grains, fractal dust aggregates, heating by
diffuse non-ionizing UV radiation, spinning dust, changes in grain
composition etc.  We believe, for various reasons, that the most
likely explanation is the erosion of large dust particles changing the
balance between big grains in thermal equilibrium and stochastically
heated very small grains, leading to a preponderance of the latter.
The RJ excess then represents the nonthermal emission from the cold
tail of the temperature distribution of the small grains.

\begin{figure*}
\begin{center}
\begin{minipage}{3.5cm}
\resizebox{3.5cm}{!}{\rotatebox{0}{\includegraphics*{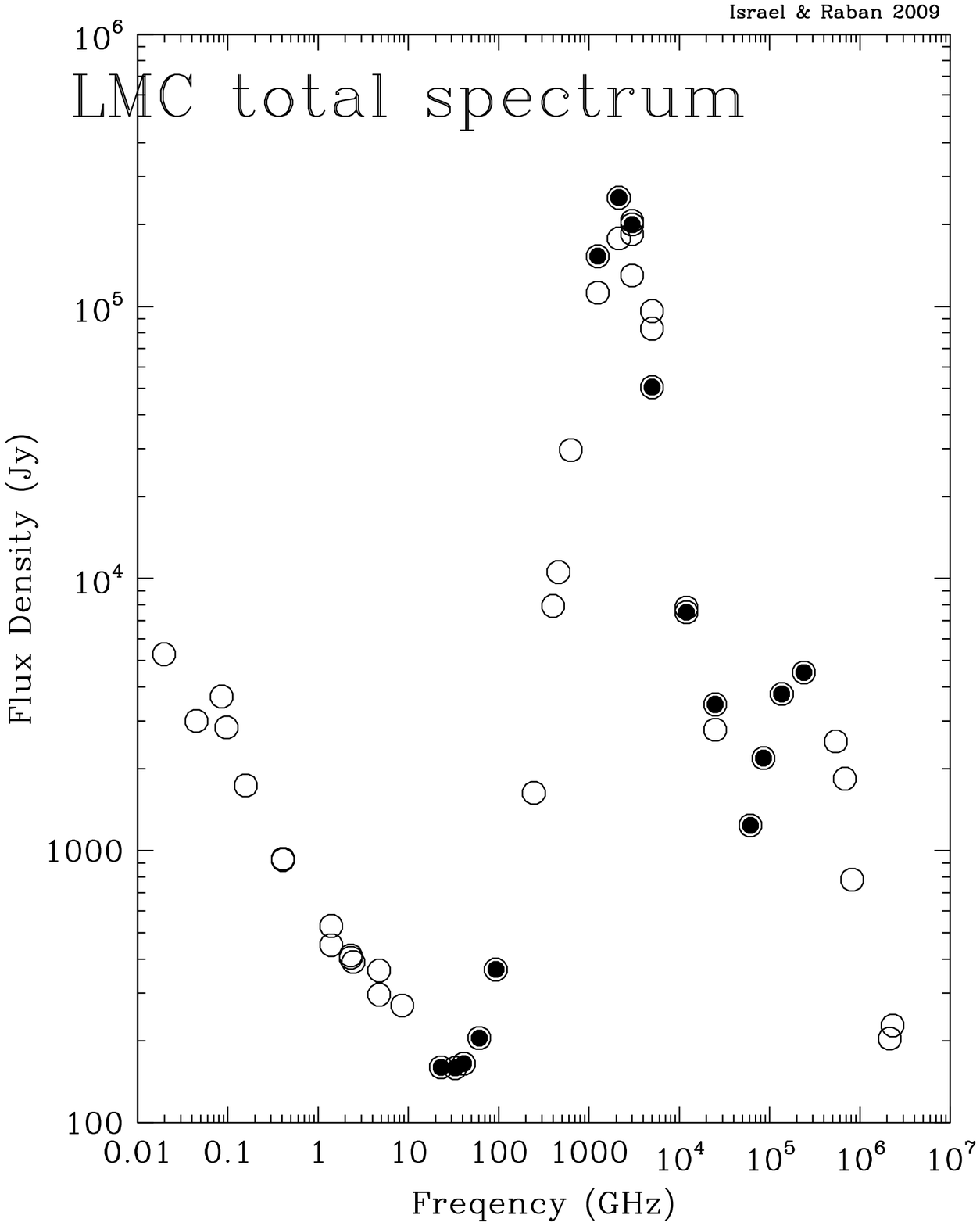}}}
\end{minipage}
\begin{minipage}{3.5cm}
\resizebox{3.5cm}{!}{\rotatebox{0}{\includegraphics*{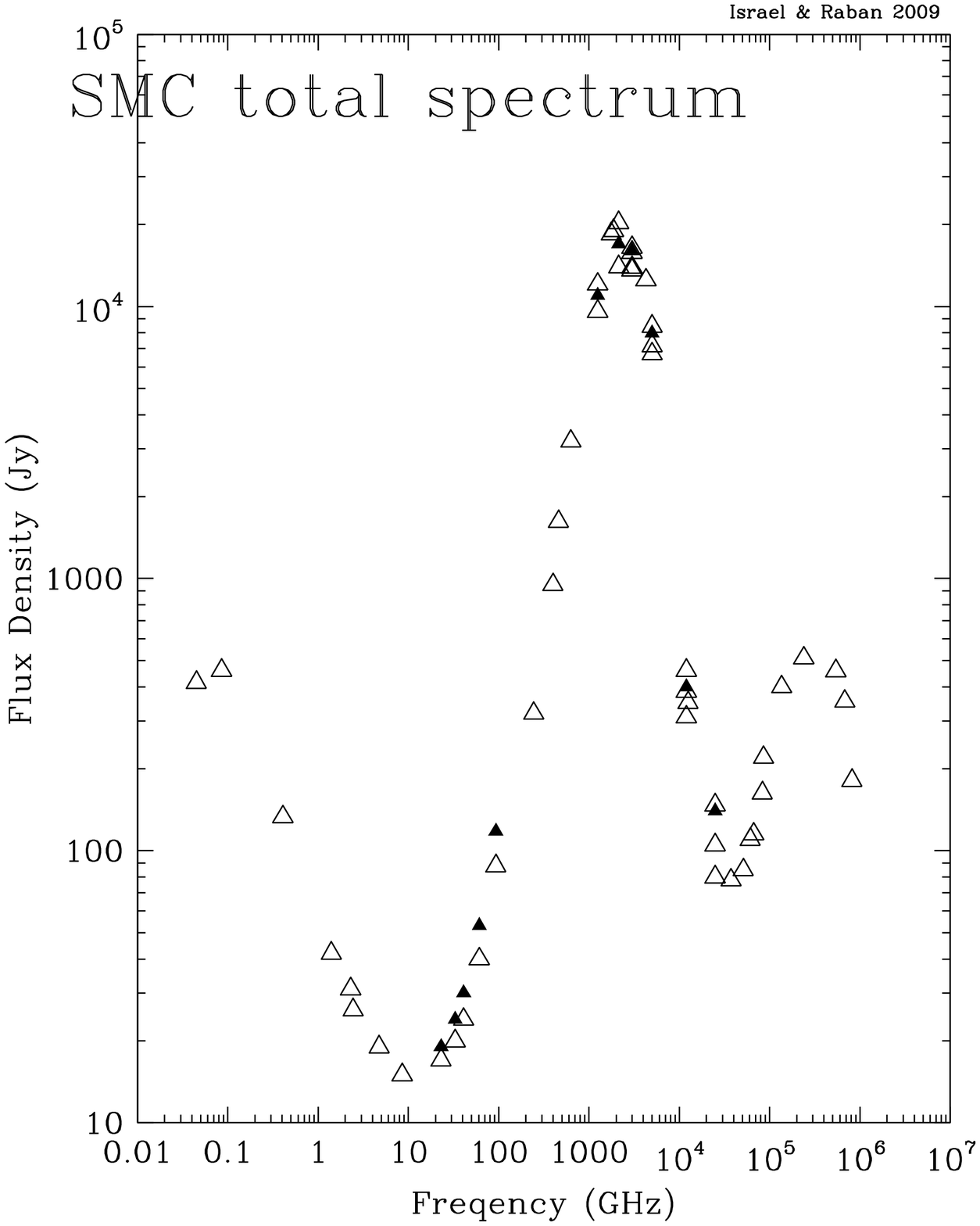}}}
\end{minipage}
\begin{minipage}{3.5cm}
\resizebox{3.5cm}{!}{\rotatebox{0}{\includegraphics*{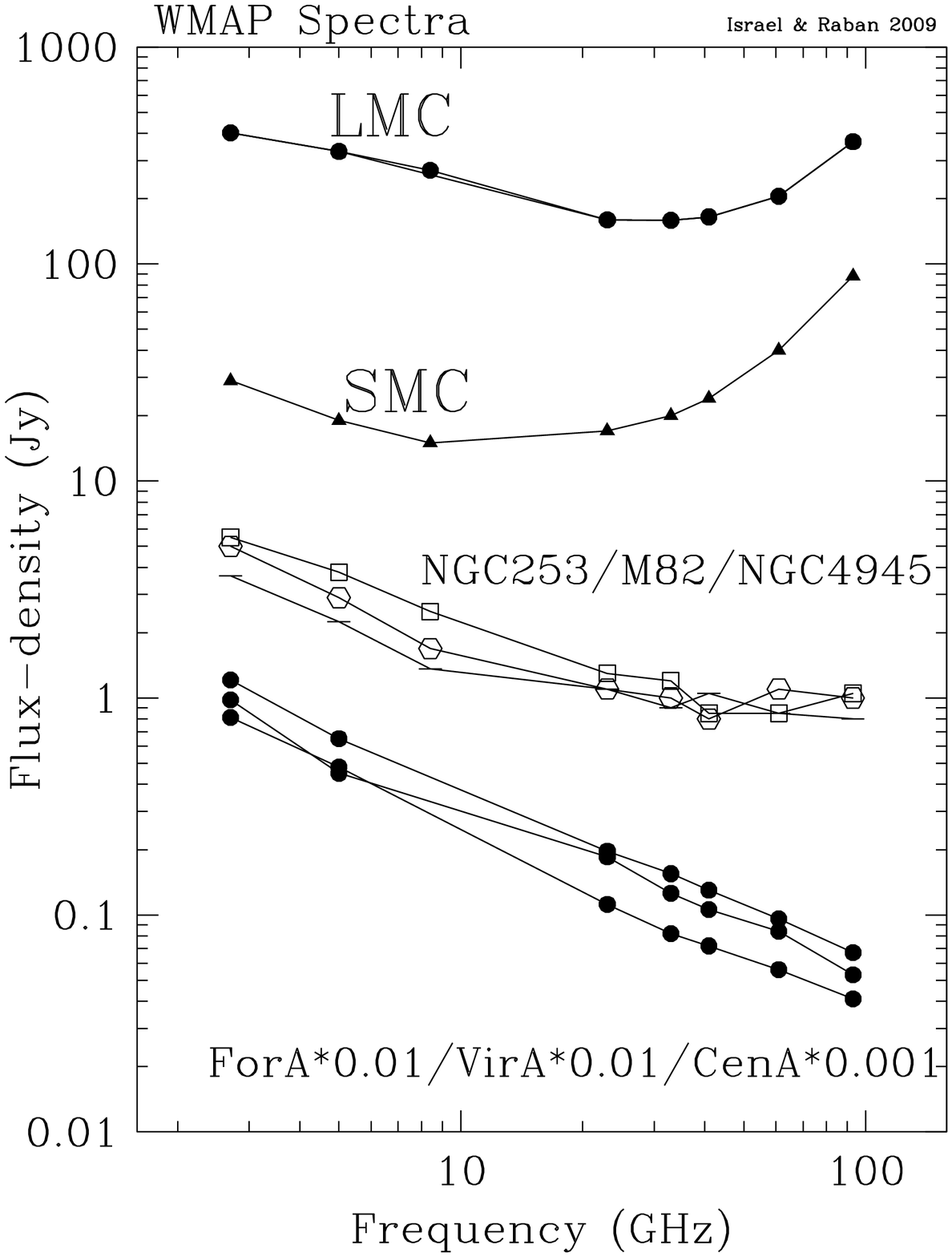}}}
\end{minipage}
\caption{In the LMC (left) and SMC (center) spectra, ranging from
  low-frequency radio range to the ultraviolet, the presence of
  Rayleigh-Jeans excess emission in the 30-1000 GHz range is clearly
  visible.  When we compare emission of these galaxies in the 10-100
  GHz range (right) to that of strong extragalactic radio sources
  (nonthermal spectrum throughout) and starburst galaxies (mostly
  thermal spectra), the magnitude of the RJ excess in the Magellanic Clouds
is even more pronounced (Israel et al. 2009).
}
\end{center}
\end{figure*}

A difficulty in interpreting the local RJ excess emission is our lack
of absolute sub-mm measurements within the Milky Way, and an even
greater difficulty in extragalactic studies is the lack of spatial
(linear) resolution and hence wide range of temperatures within a
single beam. Detailed observations of the Magellanic Clouds in the
far-infrared, sub-millimeter range would address this problem.
Specifically, it would allow us to (i) obtain accurate dust mass
estimates, (ii) provide us in detail with the dust response to various
local radiation field intensities, (iii) reveal the metallicity
dependence of these dust responses, and (iv) establish the relation
between dust processing, and molecular gas photo-dissociation.

\section{A proposed FIR/Sub-mm observing program}

At present, no systematic far-infrared/sub-millimeter survey of either
LMC or SMC exists. Short-wards of 200 microns, IRAS and Spitzer maps
are available, not allowing, however, to investigate the crucial
sub-millimeter excess.  This may become possible, for the first time, if
the Herschel Space Observatory mission is successful.  One of its
accepted Open Time Key Programmes aims at conducting uniform surveys
of the LMC and the SMC in the 250, 350 and 500 micron bands.

However, the limited aperture of the Herschel Space Observatory allows
for very big improvement - in fact essential to solving the problems
sketched above - by using the best possible site on Earth to conduct
ground-based surveys of the Clouds in the 200, 350, and 450 micron
bands with a large 25 m class telescope.  Such a telescope would be
extremely sensitive, with a 50 times larger collecting, and it would
have a linear resolution 7 times higher.  With such resolutions and
sensitivities, we will be able to systematically sample interstellar
conditions on scales of 1-2 parsec.  These scales, that correspond the
spheres of influence of individual stars, are completely out of reach
of the Spitzer and Herschel surveys.  The dust maps will show the
distribution of (cool) dust on linear scales corresponding to typical
gas column densities of the order of $10^{21}$ H-atoms/cm$^{2}$, the
threshold for the phase change of atomic into molecular hydrogen.

A similar argument can be presented in favour of sub-millimeter
spectroscopy.  Under moderately good daytime conditions, 440-900 GHz
frequency atmospheric opacities allow straightforward mapping of the
higher $^{12}$CO and $^{13}$CO transitions ($J$=4-3 to $J$=8-7), the
only two relevant C$^{\circ}$ transitions, and e.g. the $J$=5-4
and $J$=8-7 transitions of HCN, HNC, and HCO$^{+}$ that together allow
us to model the moderately warm dense and very dense molecular gas.
Under exceptional conditions, even the very important C$^{+}$ line at
9000 GHz (157 $\mu$m) may be accessible.

The resulting images will provide unique insight in the conditions
dominating dust in the interstellar medium of galaxies.  They will be
relatively little affected by large temperature-related uncertainties,
and allow study of the coldest ISM dust in the Magellanic Clouds.  The
high resolution also will provide even much better identification of
embedded young stellar objects (YSOs). The variation in dust
properties gleaned from the dust maps together with the inventory of
radiation sources and gas densities obtained from other already
existing data-sets will allow us, in particular, to study changes in
interstellar grain excitation and size distribution, i.e. follow in
detail the process of dust particle erosion in areas dominated by
radiation, by shocks, and at significantly different metallicities.

It will not be necessary to survey the full extent of the Magellanic
Clouds.  We may use the lower resolution, less sensitive Spitzer and
Herschel surveys to select specific emission regions, and specific
sectors of the Clouds for more detailed and more sensitive imaging.
Depending on the scope of such a program, emission in both LMC and SMC
may then be probed an order of magnitude deeper than in these maps.
This is important, because the observations will then reach low-mass,
low-surface column density dust and gas representative of the diffuse
ISM, which will not be reached by the Herschel results. This diffuse
ISM may be the most susceptible to the processing effects we intend to
study.

Such an observing program benefits not only from the superb quality of
the Antarctic site, but also from the fact that the Magellanic Clouds
are circumpolar, and {\it always)} high in the sky: round-the-clock
observing is quite feasible. 

The first part of a dedicated Magellanic Cloud observing program,
detailed and full modelling of dust, should be based on the
availability of large, many-pixel bolometer arrays allowing
simultaneous multiband mapping.  With only a modest extrapolation of
present-day capabilities we estimate that continuum mapping of
essential Cloud sections can be completed in a year's time, assuming
12hr/day observing. The second part, full and detailed modelling of
warm, moderately dense molecular gas using observations of C, CO, and
its isotopes, also must be based on the availability of large,
many-pixel hetero-dyne arrays operating at various frequencies. Again
assuming 12 hrs/day observing, this part requires 2-4 years for
completion depending on the scope of the program and the technical
capabilities of the then available arrays.  The results of this part
of the program can be complemented and completed with a full detailed
modelling of the subset of warm and very dense clouds, using the more
complex molecules as tracers in 1-2 years additional time.  Thus, we
envision use of a large sub-millimeter telescope on the best Antarctic
site to carry out a program to fully characterize the Magellanic
Clouds interstellar medium constitution and energetics


\end{document}